Letter

https://doi.org/10.1038/s41563-023-01483-7

# Dynamic crystallography reveals spontaneous anisotropy in cubic GeTe

Simon A. J. Kimber [1] ✉, Jiayong Zhang[2], Charles H. Liang[3,4],
Gian G. Guzmán-Verri [5,6], Peter B. Littlewood[3,7], Yongqiang Cheng[2],
Douglas L. Abernathy [2], Jessica M. Hudspeth[8], Zhong-Zhen Luo[9],
Mercouri G. Kanatzidis [9], Tapan Chatterji[10], Anibal J. Ramirez-Cuesta [2] ✉ &
Simon J. L. Billinge [11,12] ✉

Cubic energy materials such as thermoelectrics or hybrid perovskite materials are often understood to be highly disordered[1,2]. In GeTe and related IV–VI compounds, this is thought to provide the low thermal conductivities needed for thermoelectric applications[1]. Since conventional crystallography cannot distinguish between static disorder and atomic motions, we develop the energy-resolved variable-shutter pair distribution function technique. This collects structural snapshots with varying exposure times, on timescales relevant for atomic motions. In disagreement with previous interpretations[3–5], we find the time-averaged structure of GeTe to be crystalline at all temperatures, but with anisotropic anharmonic dynamics at higher temperatures that resemble static disorder at fast shutter speeds, with correlated ferroelectric fluctuations along the $<100>_c$ direction. We show that this anisotropy naturally emerges from a Ginzburg–Landau model that couples polarization fluctuations through long-range elastic interactions[6]. By accessing time-dependent atomic correlations in energy materials, we resolve the long-standing disagreement between local and average structure probes[1,7–9] and show that spontaneous anisotropy is ubiquitous in cubic IV–VI materials.

The apparently simple IV–VI material GeTe hosts unexpected electronic properties. These include thermoelectricity[10], ultrafast phase changes[11], and coupling between ferroelectric and spin degrees of freedom[12]. Underlying all these is a delicate interplay between two factors[13]: resonant $p$-electron bonding, which favours a high-symmetry cubic structure (c-GeTe), and a Peierls-type instability of the resulting half-filled band structure (Fig. 1a). The latter favours alternating long and short Ge–Te bonds, as well as rhombohedral symmetry (r-GeTe)[13–15]. Cubic GeTe (stable above $T_C$ 650 K) is attractive for device applications, because it is anharmonic and highly polarizable. However, the nature of its phase transition and even the structure of pristine c-GeTe are disputed. Local structure probes claim symmetry-breaking disorder[3–5], in disagreement with the results of spectroscopy and diffraction[16–18]. There are similar debates for other IV–VI materials[1,7–9], ferroelectrics[19,20] and hybrid perovskite solar materials[2]. We argue here that the links among structure, fluctuations and properties of high-symmetry energy materials are, in general, poorly understood.

[1]Université Bourgogne Franche-Comté, Université de Bourgogne, Nanosciences Department, ICB-Laboratoire Interdisciplinaire Carnot de Bourgogne, Bâtiment Sciences Mirande, Dijon, France. [2]Neutron Scattering Division, Oak Ridge National Laboratory, Oak Ridge, TN, USA. [3]James Franck Institute, University of Chicago, Chicago, IL, USA. [4]Pritzker School of Molecular Engineering, University of Chicago, Chicago, IL, USA. [5]Centro de Investigación en Ciencia e Ingeniería de Materiales (CICIMA), Universidad de Costa Rica, San José, Costa Rica. [6]Escuela de Física, Universidad de Costa Rica, San José, Costa Rica. [7]Materials Science Division, Argonne National Laboratory, Argonne, IL, USA. [8]ESRF, The European Synchrotron, Grenoble, France. [9]Department of Chemistry, Northwestern University, Evanston, IL, USA. [10]Institut Laue-Langevin, Grenoble, France. [11]Condensed Matter Physics & Materials Science Department, Brookhaven National Laboratory, Upton, NY, USA. [12]Department of Applied Physics and Applied Mathematics, Columbia University, New York, NY, USA. ✉e-mail: sajkimber@protonmail.com; ramirezcueaj@ornl.gov; sb2896@columbia.edu

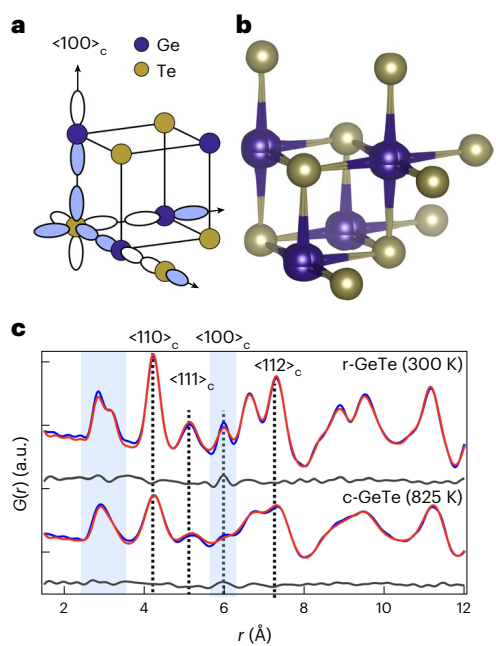

**Fig. 1 | Local distortion in GeTe and fits to X-ray PDFs in the $R3m$ and $Fm\bar{3}m$ phases. a**, Electronic structure of IV–VI materials consists of orthogonal one-dimensional bands made up of valence $p$ orbitals. These are susceptible to Peierls distortions, resolved in GeTe by a <111> shift in the Ge sublattice. **b**, Apparent disorder in c-GeTe; note the splitting of the purple Ge sites. **c**, Fit to the room-temperature PDF of GeTe using the $R3m$ structure (goodness-of-fit $R_w = 0.087$) (top). Peaks corresponding to several important distances are highlighted. The best fit of our split-site model for c-GeTe at 825 K ($R_w = 0.104$) (bottom).

Our X-ray scattering measurements highlight this issue for GeTe. Diffraction confirms the cubic NaCl structure (Supplementary Section 1) above ~650 K, and shows that c-GeTe is nearly as crystalline as silicon (Supplementary Section 2). However, fitting the nearest-neighbour peak in our X-ray pair distribution functions (xPDFs) requires local symmetry breaking, with disordered Ge positions in c-GeTe (Fig. 1b). This peak remains asymmetric to the highest temperatures measured (898 K; Fig. 1c), and has been interpreted as a 'memory' of <111>$_c$ Ge displacements in the ambient-temperature r-GeTe phase[3–5]. Curiously, we also noticed that the Ge–Ge/Te–Te peak at 6 Å is anomalously sharp at all temperatures (Fig. 1a).

Since standard crystallography cannot distinguish static disorder from dynamic motions, we developed the variable-shutter pair distribution function (vsPDF) method. This uses a time-of-flight neutron spectrometer to generate dynamic pair distribution functions (PDFs)[21] from two-dimensional datasets in reciprocal space and energy (Supplementary Section 3). In analogy with photography, we change the 'shutter speed' by varying the energy integration window ($0 < E_{max} < \infty$), and develop a principle component analysis (PCA) to separate the elastic and inelastic components of the signal[22]. Our method generates PDFs that interpolate between the time-averaged PDF, $G(r, \tau = \infty)$, and the instantaneous snapshot PDF, $G(r, \tau = 0)$.

The data collected for c-GeTe at 720 K are shown in Fig. 2, revealing how varying the energy integration window freezes, or blurs out, structural details. The instantaneous PDF, calculated by integrating to $E_{max} = \infty$, captures fast motions (Fig. 2a). Attempting to fit the average NaCl structure reveals the characteristic split Ge–Te peak at ~3 Å, and anomalously sharp <100>$_c$ peak at 6 Å seen in the xPDF data.

In contrast, the time-averaged (or elastic) PDF extracted by PCA is very well fit by the same structure model in all respects (Fig. 2b). Note that the peaks in $G(r, \tau = \infty)$ are broader at low $r$, as they represent the average positions of all the atomic configurations sampled at 720 K. This directly shows that the apparent symmetry breaking in c-GeTe simply arises from the correlated motion of atoms, which manifests as peak sharpening at low $r$. The vsPDF method, thus, reconciles the local[3–5] and average structure[16–18] of GeTe in a single measurement, showing that dynamic displacements mimic disorder in an otherwise perfectly ordered host.

This interpretation is confirmed by the pre-Fourier-transform reciprocal-space structure factors (Fig. 2c). The $E_{max} = \infty$ structure factor contains a clear high-$Q$ inelastic oscillation. This can be parameterized as ~$\sin(Q \times r)$ with $r \approx 2.88$ Å, which is the short Ge–Te distance in Peierls-distorted r-GeTe, as well as the predicted high-$Q$ limit for the multiphonon cross-section[23]. This oscillation directly corresponds to peak sharpening at 2.88 Å in real space, and is also seen in the energy-integrated X-ray scattering structure factors (Supplementary Section 1). In contrast, the $E_{max} = 0$ (or elastic) structure factor, $Q \times [S(Q) - 1]$, is almost flat at high $Q$. The first coordination sphere distortions (and <100>$_c$ sharpening at 6 Å) are thus proven to reflect dynamical correlated motion. Finally, the success of our PCA analysis is reflected in the similarity between the extracted inelastic component and fit residual (Fig. 2a). Further analysis with varying $E_{max}$ shows (Supplementary Section 4) that the crossover between time-averaged and instantaneous structures occurs at ~6 meV.

Next, we build a consistent picture of diffuse scattering, atomic motion and phonon dynamics using ab initio molecular dynamics (MD) simulations. These were performed at 720 K, and as reported previously[5,18], the X-ray weighted radial distribution function (Supplementary Section 5) reproduces the asymmetry seen in the experiment. The phonon dispersion, thermal diffuse scattering (TDS) and phonon density of states (PDOS) were then extracted in the harmonic limit[24,25]. Excellent agreement between the simulated and measured TDS is found at 720 K (Fig. 3a); however, the high-$Q$ oscillation is missing. The corresponding phonon dispersion is shown in Fig. 3b, and we find c-GeTe to be dynamically stable, with a zone centre energy of 7 meV for the $\Gamma_4^-$ soft mode. This energy scale matches our experiments (Supplementary Section 4), and the eigenvalues of this mode[17] also match the <111>$_c$ anisotropic motion identified by our X-ray scattering results.

In other IV–VI rock-salt materials, the effect of temperature on PDOS is fairly weak. By contrast, r-GeTe shows large changes on heating[18], reflecting the competition between Peierls-distorted and resonant bonded ground states[13–15]. Our data (Fig. 3c) shows pronounced softening and the emergence (Supplementary Section 6) of a peak at 11.5 meV. The position of this feature matches the calculated optic DOS; however, this (harmonic) calculation produces a very sharp peak due to the lack of lifetime broadening. In contrast, extracting the incoherent PDOS using the velocity autocorrelation function not only reproduces the peak position but also the energy envelope of PDOS. Since the energy-integrated MD simulation accounts for the diffuse background underneath the Bragg peaks (Fig. 3a), this shows that the disorder in c-GeTe is the result of anharmonic optical modes mainly of the Ge character, with strongly reduced lifetimes. Since the associated atomic fluctuations are ferroelectric, this explains the giant enhancement[26] in dielectric constant, as well as implies that the dynamics are highly anisotropic.

Returning now to our xPDF measurements, we note that these measurements also capture the total diffuse scattering from dynamic displacements (Supplementary Section 7). In fact, all the deviations from the average structure are due to such motions. Furthermore, in the low-$r$ region, the PDF contains anisotropic information, due to the high symmetry and limited peak overlap. We, therefore, separated out the dynamics by fitting the high-$r$ region (20–50 Å) of the xPDF data with the average structures determined by diffraction. The resulting misfits at low $r$ are considerable, reflecting the softness and anharmonicity of GeTe (Fig. 4a).

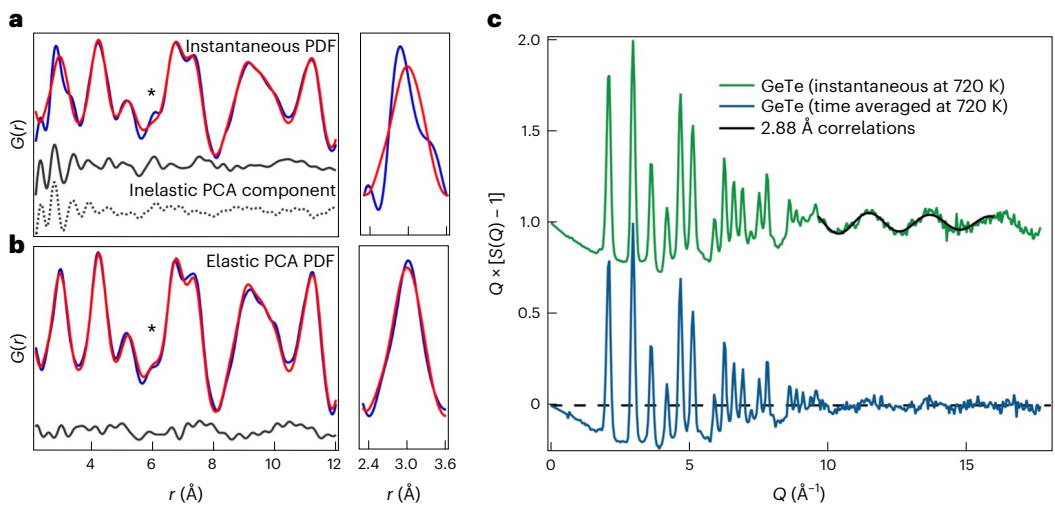

**Fig. 2 | Instantaneous and time-averaged neutron scattering results for c-GeTe at 720 K using 300 meV neutrons. a**, Instantaneous (total) PDF fitted with the average rock-salt c-GeTe structure. Obvious peak splittings and sharpening are found at low $r$. The first coordination shell splitting is shown in the inset, and the purely inelastic PDF extracted by the PCA analysis is shown to replicate the misfits between the average and instantaneous structures. The asterisk highlights the $<100>_c$ peak. **b**, Elastic PDF extracted using the PCA analysis, also showing a fit to the average NaCl structure. The peak splitting of the first coordination shell and $<100>_c$ sharpening are completely absent. The asterisk highlights the $<100>_c$ peak. **c**, Structure factors, $Q \times [S(Q) - 1]$, determined for the total (instantaneous) and elastic (time-averaged) scattering of GeTe at 720 K using ARCS. A significant extra oscillation is present in the total integrated structure factor, showing that the 2.88 Å real-space splitting is dynamic.

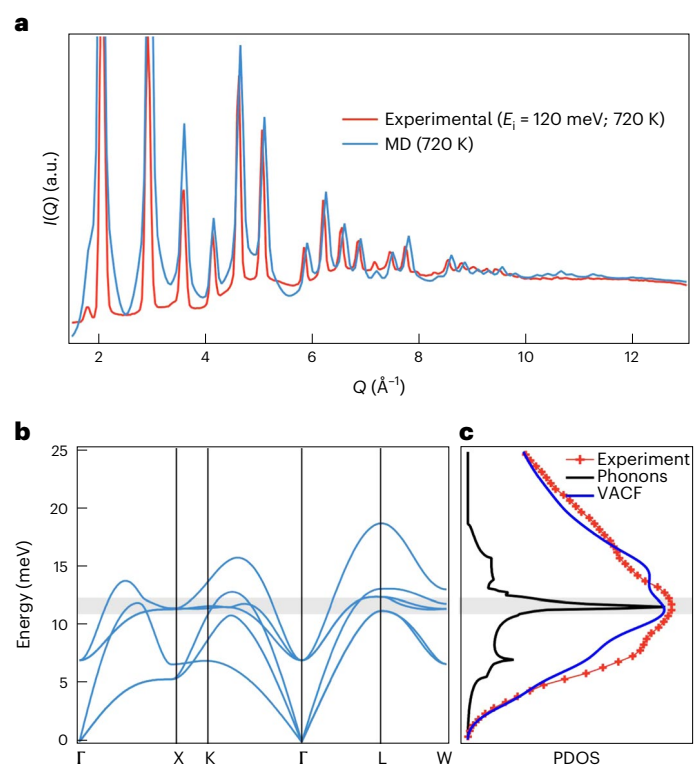

**Fig. 3 | Calculated TDS and phonon dispersion for c-GeTe. a**, Comparison of the energy-integrated ($E_i$ = 120 meV) scattering and simulation at 720 K. Note the excellent agreement of the background TDS intensity. **b**, Phonon dispersion extracted from ab initio MD simulations. In contrast to $T$ = 0 K DFT[29], the structure is found to be dynamically stable. **c**, Observed and calculated PDOS for c-GeTe at 720 K. The results are shown for a harmonic calculation from phonon dispersion and directly from the MD simulation using the velocity autocorrelation function (VACF).

The temperature dependence of the residual signal is a model-independent measure of real-space dynamics in GeTe. Plotting the peak heights as a function of temperature (Fig. 4b) reveals a clear and unexpected directional anisotropy. Although the PDF is (in principle) sensitive to elastic anisotropy[27], these effects are normally tiny, and ignored in standard refinements of structural models. As described in Supplementary Section 8, this result explains why the $<100>_c$ PDF peak is not correctly fitted for GeTe in Fig. 1a, as well as in reports for related materials in the literature. Correlations between displacements in the $<111>_c$ direction begin to fall well before the rhombohedral–cubic phase transition, and are swamped by random displacements in the cubic phase. This drastic softening is consistent with DFT calculations, which show that the Ge–Te bond order is only 0.5 (ref. [15]) in c-GeTe. In contrast, correlations along $<100>_c$ stiffen approaching the phase transition, becoming nearly temperature independent at $T > T_C$. This surprising result prompted us to develop a generic model for cubic ferroelectrics. This couples polarization fluctuations to elastic strains. The dominant coupling is to rhombohedral shear, and we imposed an elastic compatibility condition[6]. This generates an anisotropic and long-range interaction (not found in classic Landau theory) between polarization fluctuations, mediated by the strain degrees of freedom (Fig. 4c and Supplementary Section 9). The local physics favours disorder, whereas the elastic interactions produce anisotropic couplings that suppress fluctuations but only in certain directions. A two-dimensional illustration in real space is shown in Fig. 4c. Ferroelastic shear distortions can be seen to preferentially propagate in the $<10>$ direction, rather than along the $<11>$ diagonal. This model explains the counterintuitive properties of thermoelectrics like GeTe, where the effect of instantaneous disorder on electrical and thermal conductivities is rather decoupled[1]. As shown in Fig. 1a, the stiff $<100>_c$ direction corresponds to the direction of maximum $p$-orbital overlap, which forms the conduction bands[23]. Meanwhile, a snapshot of the structure along $<111>_c$ reveals instantaneous disorder, with dynamic correlations over no more than 1–2 unit cells. This length scale is optimum for scattering heat-carrying acoustic phonons, and is reminiscent of the ferroelectric large polaron proposed in lead halide perovskites[28]. To briefly summarize our results so far, we have shown the following: GeTe is perfectly crystalline at all temperatures, in disagreement with previous reports of disorder; local dynamics are dominated by a highly anharmonic correlated motion of the nearest neighbours of Ge–Te; through ferroelastic coupling, this generates long-range strains, which strengthen correlations along <100>$_c$. But how generic are our observations? First, our results explain the empirical correlation between high symmetry and performance in energy materials[1,2,10,11]. This arises because spontaneous elastic anisotropy is naturally stronger in cubic systems, as long-range strains are rapidly quenched by crystallographic degrees of freedom/flexibility[6]. High-symmetry materials are, hence, more likely to host the highly anisotropic fluctuations that enhance transport properties as described above. Second, by examining previously published PDF data, we were able to show that misfits of the <100>$_c$ peak (which indicates unusual elastic anisotropy) are ubiquitous for IV–VI and related materials (Supplementary Section 8). This effect is also seen for PbTe (ref. [1]) and SnTe (ref. [8]). According to theoretical predictions[13,14] for IV–VI materials, the fourth nearest-neighbour <100>$_c$ coupling should be activated by a large Born effective charge ($Z^*$). Indeed, the sharpening is notably weaker[29] in PbSe and absent[1] in PbS, confirming the scaling with $Z^*$.

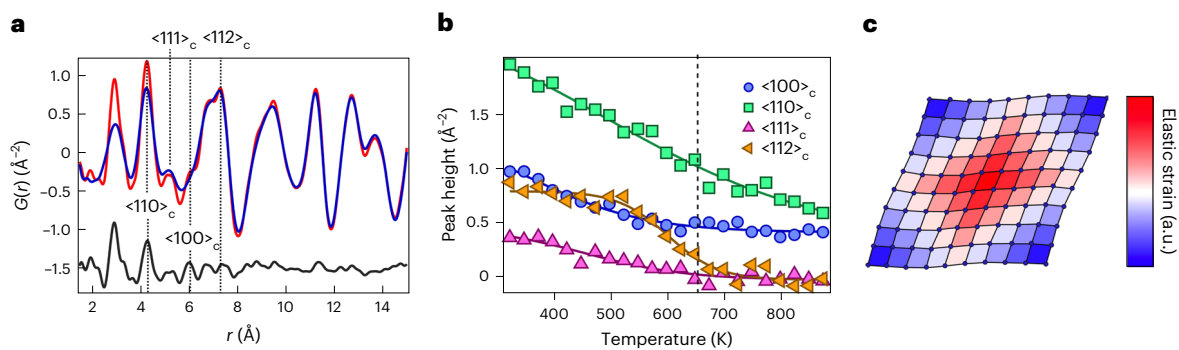

**Fig. 4 | Temperature dependence of real-space dynamics and emergence of strain in GeTe. a**, Real-space TDS for c-GeTe at 898 K. This signal (which appears in the residual) was isolated by fitting the average structure to the xPDF data in 20 < $r$ < 50 Å, and calculating $G_{obs}(r) - G_{calc}(r)$ over the $r$ range shown. **b**, Temperature dependence of the anisotropic correlated motion in GeTe. The plot shows the peak height of the selected features in the residual ($G(r) - G_{obs}(r)$); lines are guides to the eye. The data are normalized by the fitted scale factor at each temperature. **c**, Schematic of the real-space anisotropy that arises from our model of coupled ferroelectric fluctuations and shear strains. The coupling along <10> is enhanced whereas those along <11> are reduced in this two-dimensional schematic.

Similar behaviour is seen in other cubic energy materials. In $ABO_3$ perovskites, the A cation sits in a flexible cage formed by corner-sharing $BO_6$ octahedra. Anharmonic fluctuations are, thus, expected at higher temperatures. Indeed, the major feature of instantaneous PDFs of paraelectric c-$KNbO_3$ is <100>$_c$ peak sharpening[30]. Exactly the same characteristic is found for photovoltaic hybrid perovskites[2], where anomalously low resistance to <111>$_c$ shear strain has already been linked to fluctuations of the A cation[31]. In summary, GeTe (which shows important thermoelectric, phase change and electronic properties) is highly crystalline at all temperatures. Previous reports[3–5] of static, multiwell disorder are shown to be highly local, directional and anharmonic phonons. These resemble static distortions when frozen with sufficiently fast shutter speeds. We show how these motions naturally induce anisotropic elastic strains that couple them.

Very similar phenomenologies are seen in a broad range of energy materials, which include apparent local disorder[1,2,8], a central relaxational mode[32] and evidence for anisotropy[31]. We, therefore, speculate that coupling between fluctuations and strain opens up new routes to control the optoelectronic properties. We conclude by acknowledging the historical importance of measuring vibrational frequencies using neutrons[33]. These approaches are the most relevant for well-defined phonons with long lifetimes. The vsPDF approach complements such studies for cases where phonons have shorter lifetimes and distortions are localized; consequently, it is more natural to directly visualize atomic displacements in real space. We anticipate that the vsPDF technique described here will become a standard tool for reconciling local and average structures in energy materials.

## Online content

Any methods, additional references, Nature Portfolio reporting summaries, source data, extended data, supplementary information, acknowledgements, peer review information; details of author contributions and competing interests; and statements of data and code availability are available at https://doi.org/10.1038/s41563-023-01483-7.

## Methods

We used two different samples of GeTe for the X-ray and neutron scattering experiments. For the X-ray scattering experiments, we used the finely ground part of a single crystal obtained from the department of H. G. von Schnering in Max Planck Stuttgart (around 1980). This is part of the same sample used for early neutron scattering experiments[16]. For the inelastic neutron scattering experiment, an ~20 g powder sample was prepared from elemental Ge and Te, using standard solid-state methods. X-ray diffraction confirmed that pure rhombohedral GeTe was formed.

We collected the X-ray scattering data on two former beamlines of the European Synchrotron Radiation Facility (namely, ID15B and ID31). On ID15B, we used an incident energy of 87 keV and a mar345 image plate. The sample was placed in a quartz capillary and heated with a hot-air blower. At each temperature point, we collected data with two sample detector distances; thus, the data optimized for both Rietveld and PDF analysis were generated under identical conditions. The data were azimuthally integrated using pyFAI[34], and converted into real space using PDFgetX3 (ref. [35]). For conversion to real space, we used a $Q$ range of $1.0 < Q < 19.5$ Å$^{-1}$ and subtracted an experimentally determined background at each temperature point. Fits to the PDFs were performed using PDFgui[36]. On ID31, we collected data in the cubic phase using the high-resolution analyser crystal stage and an incident energy of 31 keV. Rietveld analysis was performed using the GSAS–EXPGUI package[37,38].

Inelastic neutron scattering data were collected on the ARCS spectrometer at the Spallation Neutron Source, Oak Ridge National Laboratory[39]. The sample was placed in a vanadium can, and the data were collected at $E_i$ = 40, 120 and 300 meV and temperatures of 300, 550 and 720 K. We first normalized the solid angle and efficiencies of the detector to a white-beam vanadium measurement, which was performed outside the sample environment. We then used a monochromatic $E_i$ = 300 meV vanadium measurement to correct for transmission through the MICAS-III furnace[40]. We measured an empty vanadium can at 300 and 720 K for a background, and test measurements using a B$_4$C mask in place of the sample showed that scattering of the incident beam from the sample environment were efficiently removed by the radial oscillating collimator[41]. The background and absorption corrections were performed using the Paalman–Pings macro implemented in Mantid[42]. The spectrometer has two small detector gaps at a high angle. These can be removed by extrapolation in $S(Q,\omega)$ space, as their trajectories are curved, and the signal at high $Q$ is relatively flat.

To normalize the structure factors as a function of energy transfer for Fourier transformation into real space, we used two approaches. These were (1) adding a structure-less (1 – Debye–Waller) factor and (2) ad hoc PDF extraction using PDFgetN3 (ref. [43]). Both gave equivalent results. For the structure factors shown in Fig. 3, we performed a Fourier filtering in $r$ space at a distance of 1.75 Å and below and then back-transformed to reciprocal space. As described in the main text[22], we used PCA to analyse the dynamic PDFs. This was performed using Igor Pro, and we used either linear combination or varimax rotation[44] to separate the static and inelastic components.

For finite-temperature phonon calculations of the rock-salt phase, we employed the temperature-dependent effective potential method[24,45], as implemented in ALAMODE software[46]. To incorporate long-distance interaction effects, a 2 × 2 × 2 conventional cell with 64 atoms was prepared. For the sampling of displacement–force datasets and the calculations of interatomic force constants, 2,000 ab initio MD steps with a time step of 1 fs were performed by the Vienna ab initio simulation package[47,48] in the NVT ensemble at 720 K. In these calculations, the electron–ion interaction and exchange–correlation functional were described by the projector augmented wave method[49] and generalized gradient approximation[50] with Perdew–Burke–Ernzerhof parameterization[51], respectively. First-order Methfessel–Paxton scheme[52] with a smearing width of 0.05 eV was employed to integrate the total energy in the Brillouin zone with a 4 × 4 × 4 Monkhorst–Pack $k$-point grid[53]. The energy cutoff in the plane-wave functions was set to be 228 eV. Valence electron configurations in Ge and Te were $s^2p^2$ and $s^2p^4$, respectively. The inelastic neutron scattering simulation was performed using the OCLIMAX software[54]. The $S(Q,\omega)$ map was calculated using the frequencies and polarization vectors from the DFT phonon calculations. Powder averaging, coherent scattering, temperature effects and higher-order excitations (up to $n$ = 10) are included. A resolution function consistent with the experiment was applied.

## Data availability

The data underpinning this work can be found at https://doi.org/10.17605/OSF.IO/M7GXH. This includes the X-ray PDFs of GeTe as a function of temperature, the elastic and total neutron PDFs (Fig. 2), the inelastic neutron scattering data for GeTe at 720 K used to produce the dynamic PDFs and the ab initio MD trajectory for GeTe at 720 K used to extract the quantities shown in this Letter.

50. Perdew, J. P. et al. Atoms, molecules, solids, and surfaces: applications of the generalized gradient approximation for exchange and correlation. *Phys. Rev. B* **46**, 6671–6687 (1992).
51. Perdew, J. P., Burke, K. & Ernzerhof, M. Generalized gradient approximation made simple. *Phys. Rev. Lett.* **77**, 3865–3868 (1996).
52. Methfessel, M. & Paxton, A. High-precision sampling for Brillouin-zone integration in metals. *Phys. Rev. B* **40**, 3616–3621 (1989).
53. Monkhorst, H. J. & Pack, J. D. Special points for Brillouin-zone integrations. *Phys. Rev. B* **13**, 5188–5192 (1976).
54. Cheng, Y., Daemen, L., Kolesnikov, A. & Ramirez-Cuesta, A. Simulation of inelastic neutron scattering spectra using OCLIMAX. *J. Chem. Theory Comput.* **15**, 1974–1982 (2019).

## Acknowledgements
We thank the European Synchrotron Radiation Facility for the provision of beamline time on ID15B and ID31. This research used resources at the Spallation Neutron Source, a US Department of Energy (DOE), Office of Science User Facility, operated by the Oak Ridge National Laboratory. The computing and software resources were made available through the VirtuES and the ICEMAN projects, funded by the Laboratory Directed Research and Development program (LDRDs 7739, 8237,10447) and Compute and Data Environment for Science (CADES) at the Oak Ridge National Laboratory, which is supported by the Office of Science of the DOE under Contract DE-AC05-00OR22725. We thank A. Hill, R. Mills and G. Ganroth for assistance, and A. Tennant and R. Ibberson for supporting a 2017 workshop on Advanced Fourier Methods at the Shull Wollan Center. We thank T. Egami, A. Fitch, P. Senet and M. Wuttig for useful discussions. S.J.L.B. acknowledges support from the US DOE, Office of Science, Office of Basic Energy Sciences, under contract no. DE- SC0012704. C.H.L. acknowledges support from NSF GRFP DGE-1746045. G.G.G.-V. acknowledges support from the Vice-Rector for Research at the University of Costa Rica (project no. 816-C1-601). Work at Argonne (P.B.L.) is supported by the US DOE, Office of Science, Office of Basic Energy Sciences, Materials Sciences and Engineering, under contract no. DE-AC02-06CH11357. At Northwestern University (M.G.K.), work on thermoelectric materials is primarily supported by the US DOE, Office of Science, Office of Basic Energy Sciences, under award no. DE-SC0014520. This work was supported by the Programme of Investments for the Future, an ISITE-BFC project (contract no. ANR-15-IDEX-0003) (S.A.J.K.).

## Author contributions
T.C. proposed the first X-ray experiments, and provided the sample. S.A.J.K. and J.M.H. performed the X-ray experiments. S.A.J.K. and D.L.A. performed the neutron experiments. Z.-Z.L. and M.G.K. provided the neutron sample. J.Z., Y.C. and A.J.R.-C. performed and interpreted the MD simulations. C.H.L., G.G.G.-V. and P.B.L. developed the elastic compatibility theory. S.A.J.K. and S.J.L.B. developed the vsPDF analysis protocols. S.A.J.K. wrote the paper with comments from S.J.L.B. and all the other co-authors.

## Competing interests
The authors declare no competing interests.

## Additional information
**Supplementary information** The online version contains supplementary material available at https://doi.org/10.1038/s41563-023-01483-7.

**Correspondence and requests for materials** should be addressed to Simon A. J. Kimber, Anibal J. Ramirez-Cuesta or Simon J. L. Billinge.

**Peer review information** *Nature Materials* thanks the anonymous reviewers for their contribution to the peer review of this work.

**Reprints and permissions information** is available at www.nature.com/reprints.